\documentclass[aps,prl,reprint,groupedaddress,showpacs]{revtex4-1}
\usepackage{graphicx}
\usepackage{dcolumn}
\usepackage{bm}
\usepackage{color}
\graphicspath{{figures/}}
\usepackage{hyperref}
\hypersetup{backref,
pdfstartview=Fit,
colorlinks=true,
citecolor=blue,
urlcolor = blue,
linkcolor = blue}
\newcommand{\Lagr}{\mathcal{L}}
\begin{document}

\title{Enhanced DC-Field Metrology Using Nitrogen Vacancy Spins via Strong Collective Coupling and Fano-Scattering}

\author{Haitham A. R. El-Ella}
\email[]{elel@fysik.dtu.dk}
\author{Ulrik Lund Andersen}
\author{Alexander Huck}

\affiliation{Department of Physics, Technical University of Denmark, 2800 Kongens Lyngby, Denmark}

\begin{abstract}
Nitrogen-vacancy ensembles promise an unprecedented combination of sensitivity and resolution for magnetic field sensing under ambient conditions. However, their full sensing potential is limited by the inhomogeneous degradation of their collective coherence, while dynamical-decoupling limits sensing to the detection of AC-fields. Here we propose using the recently demonstrated cavity-protection effect for room-temperature, ensemble-based DC metrology. We identify a parameter space where strong-coupling can persist at room temperature and where the dressed states are limited solely by the dephasing of the cavity mode and the ensemble longitudinal relaxation, showing aT  Hz$^{-\frac{1}{2}}$ DC-field sensitivities to be theoretically achievable with further augmentation through Fano-interference.       
\end{abstract}


\maketitle

Magnetic metrology schemes employing nitrogen vacancy (NV) spin ensembles promise an exceptional combination of sensitivity and spatial resolution, due to the increased fluorescence yield and $N^{-\frac{1}{2}}$ scaling of their read-out spin-projection noise \citep{Taylor2008}, where $N$ is the number of spins in the ensemble. Due to the ease of their coherent control under ambient conditions and their bio-compatibility, NV centres have emerged as particularly attractive quantum-bits for exploring quantum information storage and transferral \citep{Maurer2012,Togan2010}, as well as sensors for studying nano-scale electromagnetic fields and temperature gradients in biological specimens \citep{LeSage2013,Dolde2011,Kucsko2013}. However, there are severe limitations with increased ensemble size based on the cumulative degradation of ensemble coherence through both intrinsic and extrinsic sources of inhomogeneity. Conventional techniques used to suppress decoherences resulting from inhomogeneity limits the overall detection bandwidth of NVs in such a way that makes DC-fields below $\sim1$ $\mu$T difficult to measure \citep{Chaudhry2014,Hall2009}.

Here we propose employing dressed spin-ensemble states for DC-magnetic field sensing. The studied scheme would involve monolithic microwave resonators with high room-temperature quality factors (\textit{Q}'s) that would facilitate optically detected magnetic resonance (ODMR). We explore the feasibility of obtaining strong spin-coupling at room temperature and the DC-sensitivity of the resulting dressed states, with further enhancement proposed through Fano-based interference with a second cavity mode. The sensing described here employs spin-resonance fluorescence-dependent detection, however phase-contrast based detection (Ramsey-type schemes) are equally applicable. Room temperature ensemble-QED and macro-scale microwave cavities have been considerably studied \citep{McNeilage2004,Resonator2013,Farr2014,Zhang2014,Tabuchi2014,Zhang2011}, and the cavities described in these references are anticipated to provide a basis for the proposals outlined in this article.

The dominant decoherence source for NV’s stems from dipole-interactions with the surrounding Overhauser field, in addition to the local variations of the crystal field resulting from free charges, strain and surface states \citep{Bar-Gill2012,Bar-Gill2013,Romach2015}. As a result, the dispersion of NV Larmor frequencies for an ensemble is inhomogeneously broadened. Furthermore, ensembles are intrinsically more sensitive to the amplitude and frequency fluctuations of control fields which poses difficulties for wide-field metrology. This is further exasperated by the random distribution of four possible crystallographic NV orientations. Protocols such as Hahn echo, CPMG-\textit{n} or XY-\textit{n} have been routinely used to extend coherence times, but result in optimised sensitivity only at bandwidths inversely proportional to half the systems total coherence time ($T_2$), and when phase-locked with an external AC-field to maximise the measured phase contrast\citep{Naydenov2010,DeLange2010,Stanwix2010}. With rigorous noise cancellation, sensitivity in the pT Hz$^{-\frac{1}{2}}$ range for AC-fields has been demonstrated using a Hahn-echo pulse scheme \citep{Wolf}.

\begin{figure*}
\centering
\includegraphics[scale = 0.9]{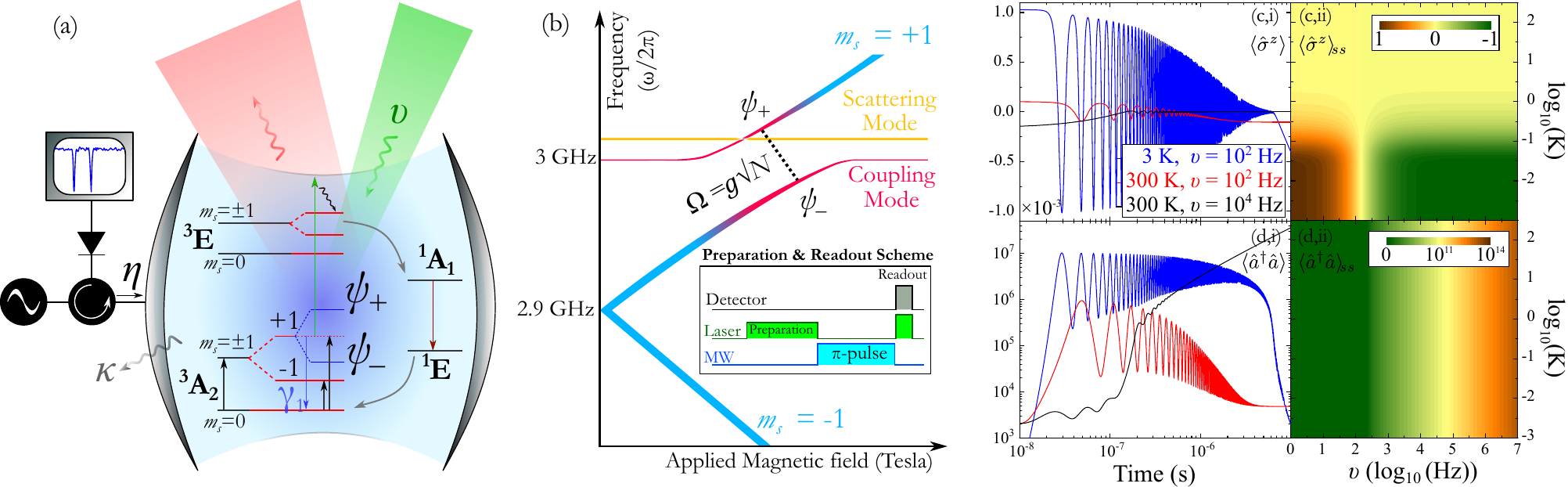}
\caption{\label.(Colour online) (a) Schematic of the cavity-ensemble system, where the dressed states $\psi_{\pm}$are accessible for an incoherent probe field with rate $v$, and the resulting fluorescence is detectable through the same or similar aperture. (b) Expected anti-crossing of the ODMR  lines when tuned into resonance with a microwave cavity mode, with the splitting being a function of the ensemble size $N$. Scattering of the dressed state can occur with nearby phase-mismatched cavity modes (see Fig.2(c,i)). Inset is the intended pulse-scheme. (c, d, i) Time-dependent decay of $\langle\hat{\sigma}^{z}\rangle$ and $\langle\hat{a}^{\dagger}\hat{a}\rangle$ for an ensemble size of $10^{20}$ with $\kappa  = 3$ kHz ( \textit{Q} = $10^6$), $\gamma_{1}= 250$ Hz, $\gamma_{2} = 1$ MHz, $g_k = 1$ Hz (corresponding to a mode volume of $~2\times10^{-3}$ cm$^{3}$ at a resonant frequency $\omega_{c}=$ 3 GHz). (c, d, ii) Steady-state values $\langle\hat{\sigma}^{z}\rangle_{ss}$ and $\langle\hat{a}^{\dagger}\hat{a}\rangle_{ss}$ plotted as a function of incoherent pump rate and temperature, starting from a thermally equilibrated state (the same parameters as in (i) are used).}
\label{Fig1}
\end{figure*}

While there is much incentive in developing optimised quantum-control and noise-discrimination protocols, alternative approaches have been explored and demonstrated in the context of hole-burning \citep{Kehayias2014}, weakly-coupled cavities masing between $m_s=0 \rightarrow-1$ states \citep{Jin2014}, strongly coupled ensemble systems \citep{Kubo2010,Sandner2012}, and recently a proposal of  combining both strong-coupling and hole-burning \citep{Krimer}. In particular, the suppression of inhomogeneous-related decoherence in strongly coupled systems, the \textit{cavity-protection} effect, has been theoretically outlined for NV spin ensembles \citep{Sandner2012} and has been recently demonstrated experimentally using a superconducting strip-line resonator \citep{Putz2014}. 

The intention here is to utilise the cavity protection effect in the scheme described in the Fig.1(a,b). Cavity-protection ensures that the  identical constituents of a spin ensemble resonant with the cavity transition (or dispersed within the cavity bandwidth) are isolated from the in-homogeneously dispersed subset, provided that the coupling strength exceeds the sum of the cavity bandwidth and the inhomogeneous dispersion width. In particular, the strong-coupling suppresses the in-homogeneously dispersed tails of the spin-distribution, effectively turning them into dark-states \citep{Diniz2011}. Such a system provides the distinct advantage of suppressing detrimental broadening caused by (coherent) control-field errors, retaining coherences similar to those of single NVs, while providing the increased fluorescence yield of an ensemble measurement when probed incoherently.   

NV spins are prepared in the ground state ($m_{s}=0$) using an incoherent optical pump with a rate $v$, which is applied for a duration that is long enough to ensure that the ground-state spin population is maximised and has reached a steady-state \citep{Harrison2004}. This is followed by a coherent microwave $\pi$-pulse with a strength $\eta$ and a frequency $\omega_{p}$ that is resonant with the cavity and a chosen spin transition (in Fig.1(b) the +1 state is chosen), which is tuned into resonance with the cavity mode by applying an external magnetic field. When resonant, the resulting dressed states $\psi_{\pm}$  are formed from a collective spin-wave (Dicke state) which posses a splitting $\Omega$ and a degree of inhomogeneous suppression that is dependent on the initial spin dispersion and number of in-phase spins. This is in turn primarily determined by the thermal bath and the initial coherent/incoherent pump rates. The dressed states are then read-out using a saturating incoherent pulse, destroying the collective spin-wave and resetting a majority of the spins into the ground-state \citep{Harrison2004}.  
    
In general, a collective spin-wave in an ensemble can be created and sustained at pump rates where the number of induced excitations for a given duration does not exceed the number of ensemble constituents, assuming all spins are identically polarised prior to excitation. In the solid-state this is difficult at elevated temperatures due to the constant exchange of energy between the lattice and the spins, yet it is still possible depending on the ensemble constituent size, collective dephasing rate and, most importantly, the amplitude of the polarising excitation. This is difficult to theoretically model in the solid state, except at low temperatures where one can justifiably collectively treat the spin ensemble under the Holstein-Primakoff (HP) approximation. This approximation allows for the formal definition of the spin operators in terms of boson operators, under the assumption that the number of excitations within a collective spin wave is much less than the number of its constituents. To justifiably apply this approximation to estimate the system output at room temperature, we are limited to theoretically explore a parameter space where the incoherent  preparation pump rate does not exceed the cavity and ensemble decay rates, as well as account for the thermal distribution of the ground state spins following the incoherent preparation pump, before the polarising coherent $\pi$-pulse. For the sake of the following derivations, we also assume the the $\pi$-pulse is weak enough to ensure that the HP-approximation holds.        

The interaction of a cavity mode with an ensemble of $N$ spins can be described by a general Hamiltonian using Pauli spin operators, and the dipole and rotating wave approximations:
\begin{equation}
\hat{\mathcal{H}} =\omega_{c}\hat{a}^{\dagger}\hat{a}  +\frac{1}{2}\sum_{k}^{N} \omega_{k}\hat{\sigma}_{k}^{z} +\sum_{k}^{N}g_{k}(\hat{\sigma}_{k}^{+}\hat{a} -\hat{\sigma}_{k}^{-}\hat{a}^{\dagger})
\end{equation}
where $\hbar=1$, $\omega_{k}$ and $g_{k}$ are the transition angular frequency and coupling strength of the $k^{th}$ spin. The dynamics which account for the cavity decay rate $\kappa$, spontaneous emission $\gamma_{1}$ ($=1/T_{1}$) and coupling to the thermal environment $\bar{n}$, polarisation dephasing $\gamma_{2}$ ($=1/T_{2}^*$), and loss through incoherent pump rate $v$ in the context of ODMR, can be numerically calculated using a master equation for a reduced density matrix: $\dot{\rho}_{s} = \frac{1}{i}[\hat{H},\rho_{s}]  +  \Lagr_{\kappa(\bar{n})}[\rho_{s}]  + \Lagr_{\gamma_{1}(\bar{n})}[\rho_{s}] +\Lagr_{\gamma_{2}}[\rho_{s}]+  \Lagr_{v}[\rho_{s}]$ (see Appendix A). Following similar justifications as in \citep{Meiser2009}, this is expanded neglecting third and higher-order correlations to explore the temperature and incoherent pump rate parameter space where a HP-based description remains valid. The relevant parameters in this case are the steady-state values for the inversion operator $\langle\hat{\sigma}^{z}\rangle_{ss}$ and the intra-cavity photon number $\langle\hat{a}^{\dagger}\hat{a}\rangle_{ss}$ prior to the application of the coherent $\pi$-pulse. It is assumed that the HP condition is fulfilled as long as $\langle\hat{a}^{\dagger}\hat{a}\rangle_{ss}$ is below the spin number $N$ in the ground state prior to polarisation into the $m_s = +1$ state, so as to allow polarisation to occur dominantly by the coherent microwave pump rather than through incoherent remnant thermal excitations. 

\begin{figure*}
\centering
\includegraphics[scale = 0.9]{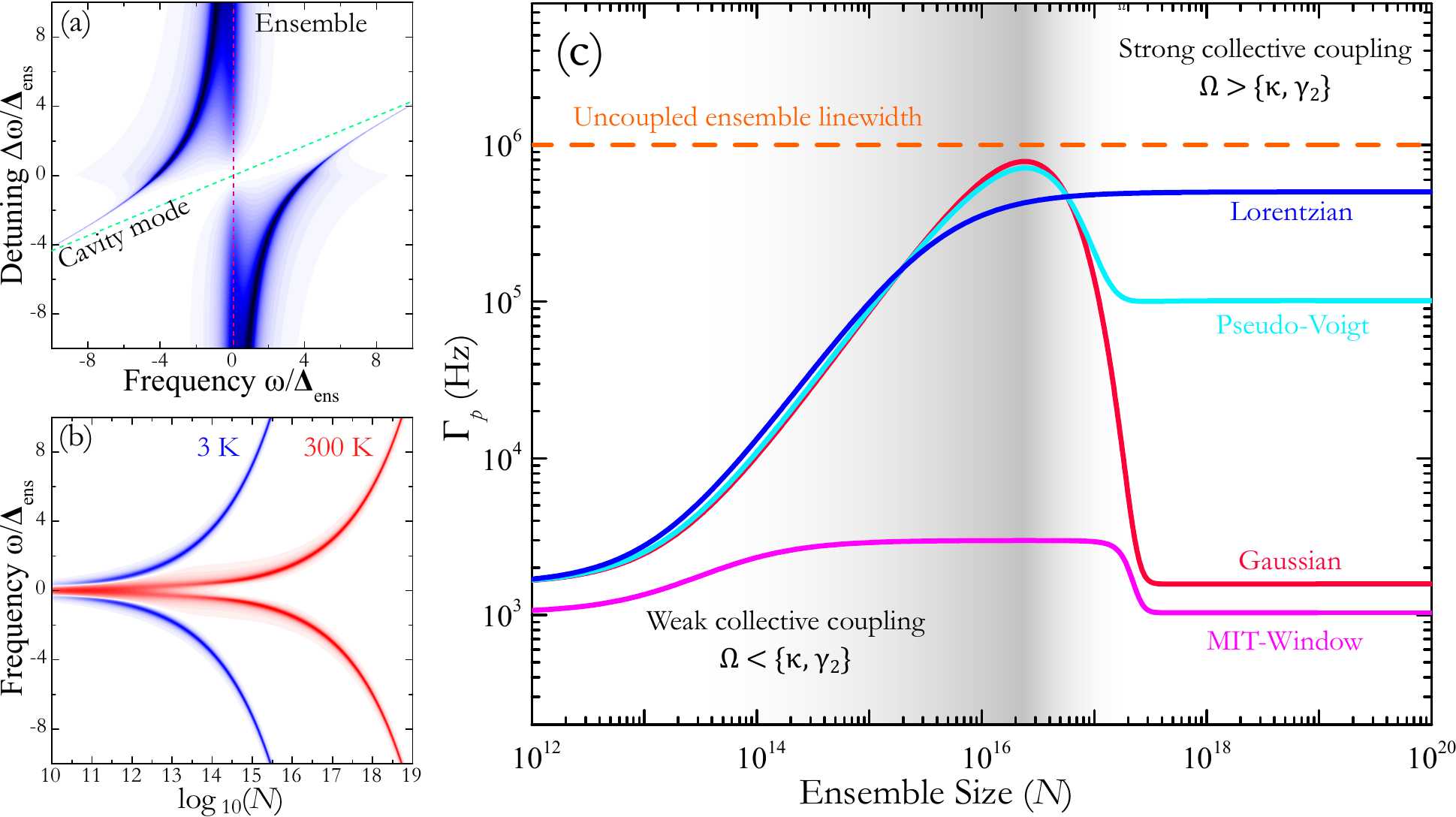}
\caption{\label.(Colour online) (a) Detuning-dependent ($\Delta\omega=\omega_{c}-\omega$) ODMR plot of a strongly coupled system with the axis divisions normalised in units of the inhomogeneous linewidth $\Delta_{en}$. (b) Surface plot of $t_{c}(\omega)$ (Eq.2) as a function of ensemble size $N$ for two different temperatures, highlighting the need for larger collective coupling strengths to generate dressed-state at higher temperature. (c) Eq.3 plotted for three different distributions (a 1:4 Psuedo-Voigt function is plotted), along with the resulting linewidth of a scattered dressed-state (Eq.4) assuming a Gaussian spin distribution. In this plot $\kappa = 3$ kHz (\textit{Q} = $10^6$), $\gamma_{1}= 250$ Hz, $\gamma_{2}=\Delta_{en}=1$ MHz,  $g = 1$ Hz, and T = 300 K.}
\label{Fig2}
\end{figure*}

As shown in \citep{Henschel2010}, the full set of coupled equations need not be solved, as the spin population will have reached a steady-state before the coherent microwave pump is reapplied. The simplified derivation results in four coupled equations and a closed pair of steady-state expressions (see Appendix A). Fig.1(c,d) shows a summary of both the temperature and incoherent pump-dependent dynamics obtained for $\langle\hat{\sigma}^{z}\rangle$ and $\langle\hat{a}^{\dagger}\hat{a}\rangle$ for an ensemble of $10^{10}$ NV’s, where the time-dependent dynamics are calculated assuming a thermally distributed spin population at $t<0$. The dynamics are presented to show the combined effect of temperature, dephasing and incoherent pump rate on the dressed state phase coherence, as well as highlighting a limit where $v$ exceeds the ensemble decay rate and destroys the coherent dressed state. As long as $v \leq\lbrace\gamma_{1}, \kappa \rbrace$,  the number of excitations accumulated at the systems steady-state prior to the application of the coherent $\pi$-pulse will be low enough to justify a HP-based description at any temperature when preparing and populating the dressed states. Given the exceptional rate of $\gamma_{1}$ for NV's \citep{Jarmola2012} such excitation rates correspond to powers in the pW-fW range depending on the system, while higher powers in the preparation stage will diminish the initial ground-state spin population and phase coherence. The steady-state values of the inversion operator also highlights the dominant effect of temperature irrespective of $v$, where the fraction of excited spins is diminished to a thermally mixed state with increasing temperatures through the relation extracted from the fit of Fig.1(c,ii): $\langle\hat{\sigma}^{z}(\text{T})\rangle_{ss}=\tanh{(\frac{\omega_{0}}{4\pi k_{b}\text{T}})}\langle\hat{\sigma}^{z}(0)\rangle_{ss}$.

Eq.1 may now be described using normalised collective operators with a temperature dependent collective coupling strength while applying the HP-approximation. Assuming all spins are equally coupled to a magnetic mode with their quantisation axis aligned with the cavity field, a cooperative coupling strength can be defined in terms of the temperature tanh-dependency extracted from the fit of Fig.1(c,ii) as a temperature-dependent (and therefore $N$-dependent) coupling strength: $\Omega_{T,N} = g\sqrt{N\tanh\omega_{0}/4\pi k_{b}T}$. Here $g\propto \gamma_{e}\sqrt{\omega_{c}/V_{\text{eff}}}$ is the individual magnetic coupling field strength defined in terms of the gyromagnetic ratio $\gamma_{e}\sim2.8$ MHz/G and the effective mode volume $V_{\text{eff}}$. By accounting for the temperature-dependant number of spins, the disordered portion of the spin ensemble is excluded, thereby only accounting for the spins that occupy the same eigenstate. Eq.1 is rewritten as: $\hat{\mathcal{H}}_{HP} =\omega_{c}\hat{a}^{\dagger}\hat{a}  + \sum_{k}\omega_{k}\hat{b}^{\dagger}\hat{b} +\sum_{k}g_{k}(\hat{b}^{\dagger}\hat{a}-\hat{b}\hat{a}^{\dagger}) $ \footnote{When $\Omega/\omega_{c}$ approaches unity, the rotating wave approximation starts to lose its validity. However, the steady-state spectrum can be calculated to first order using this approximation due to the magnitude of $g$. This is confirmed by e.g. the measured ultra-strongly coupled spectrum in \citep{Zhang2014}}, and using the input-output formalism, the steady-state cavity output $t_{c}(\omega)$ can be defined when $\gamma_{1}$ is smaller than the inhomogeneous width ($\gamma_{1}\ll\Delta_{en}=\gamma_{2}$) \citep{Diniz2011,Kurucz2011} as: 
\begin{equation}
t_{c}(\omega) = \frac{i\kappa/2}{\Delta\omega+i(\kappa+\gamma_{1}+2\pi\Omega_{T,N}^{2}\rho(\omega,\Delta_{en}))-\Omega_{T,N}^{2}\Lambda}
\end{equation} where $\Delta\omega (= \omega_{c}-\omega)$ is the detuning between the input field and the cavity resonant frequency, $\rho(\omega,\Delta_{en})$ is the spectral distribution with a full width at half maximum of $\Delta_{en}$, and $\Lambda$ is the non-linear Lamb-shift. ODMR will depopulate all ground-state levels and highlight both strongly-coupled and uncoupled spin levels. The expected spectra is plotted in Fig.2(a) for a Gaussian spin distribution, highlighting the occurrence of both cavity-pulling and suppression of emission detuned from the dressed state energies when fully on-resonance (see Appendix B for expression) . These effects are respectively diminished and enhanced as the coupling strength is increased and the dressed-state manifestation occurs at larger detuning (a small coupling strength was used to highlight these effects in Fig.2(a)). The temperature-dependent cavity output as a function of ensemble size is plotted in Fig.2(b). The dressed-states linewidth can be extracted from Eq.2 at the dressed-state frequencies ($\omega = \Omega_{T,N}$) as:
\begin{equation}
\Gamma_{p} = \frac{\kappa + \gamma_{1}+ 2\pi\Omega_{T,N}^{2}\rho(\Omega_{T,N},\Delta_{en})}{2} 
\end{equation} 

Eq.3 is plotted in Fig.2(c) for various spectral distributions using the temperature-dependent coupling strength. $\Gamma_{p}$, and thus the coherence time of the dressed-state, is directly dependent on the spectral dispersion. In the context of single NV’s where dephasing is homogeneous (Lorentzian) the influence of coupling to a cavity is detrimental. This is in contrast to ensembles which suffer from inhomogeneous broadening (the larger/denser, the more Gaussian-like), for which strong coupling will significantly suppress $\gamma_{2}$. In the limit of infinite coupling, Eq.3 is reduced to $(\kappa + \gamma_{1})/2$, for which the onset can be seen in Fig.2(c). It is worth mentioning that even for mode volumes in the order of the resonant wavelength,  the dipole-coupling between NV's ($\gamma_{e}^{2}\mu_{0}/2h\approx 3.3\times10^{-13}$ Hz  cm$^3$) will always be comparably negligible to the collective coupling strength.

\begin{figure*}
\centering
\includegraphics[scale = 0.88]{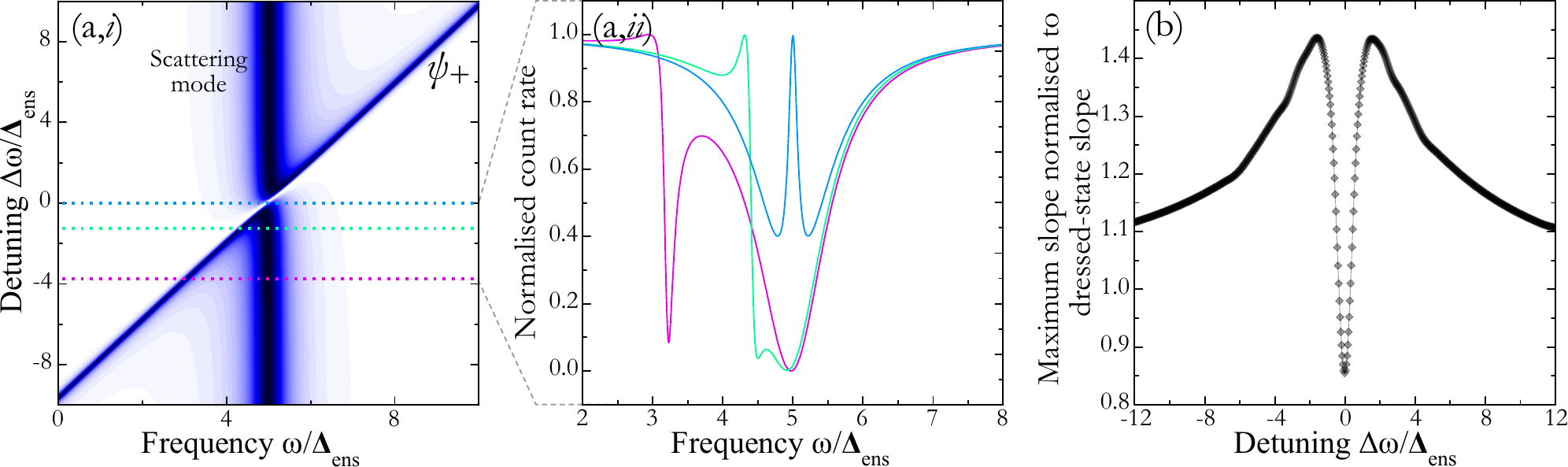}
\caption{\label.(Colour online) (a,i) Detuning-dependent ($\Delta\omega=\Omega_{T,N}-\omega_{\text{scat. mode}}$) ODMR plot of a dressed-state with a scattering mode, where the dressed-state's cavity mode quality factor is equivalent to that of the scattering mode (\textit{${Q_{p}}$} = \textit{${Q_{s}}$}). (a,ii) Plot of normalised extracted spectra from the coloured dotted lines in (a,i). (b) Plot of the maximum slope in the resulting spectra as a function of detuining, normalised to the the slope of the individual dressed-state. The maximum slope of the MIT linewidth is in the case worse then that of the un-scattered dressed state, while the parameters used are identical to those for the plots in Fig.2.}
\label{Fig3}
\end{figure*}

\begin{figure*}
\centering
\includegraphics[scale = 0.9]{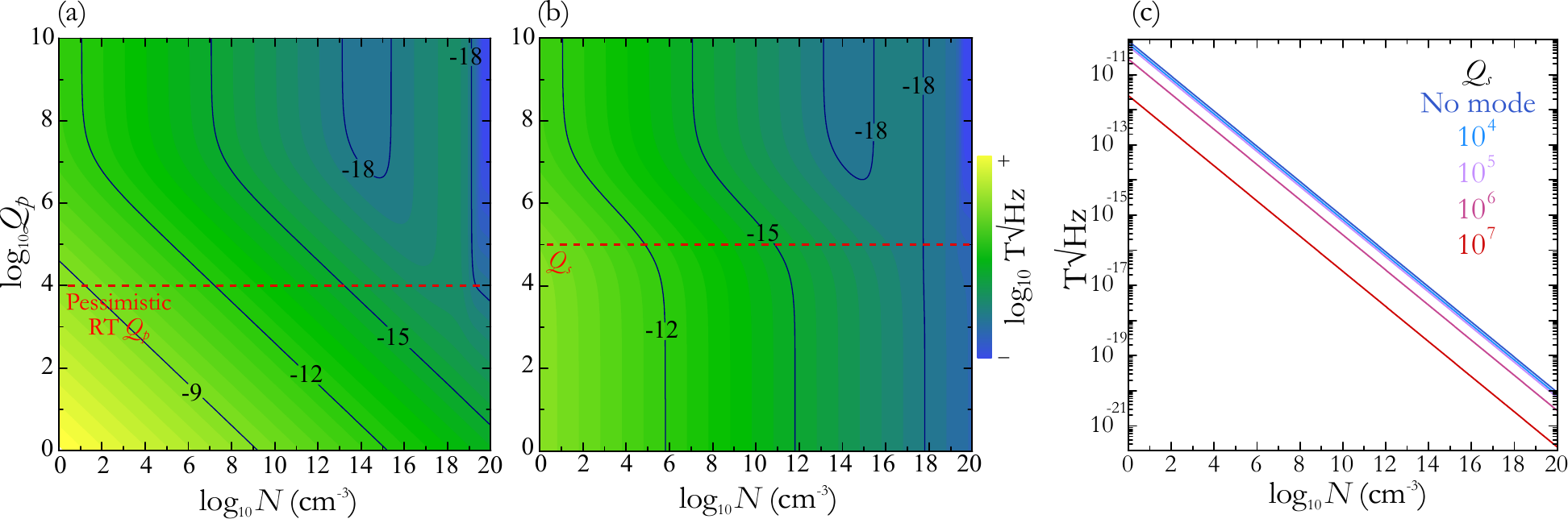}
\caption{\label.(Colour online) (a) Log-scale DC-sensitivities at 300 K as a function of \textit{$Q_{p}$} and a given density of NV’s  collectively coupling to a cavity mode with an effective mode volume of 1 cm$^{3}$ ($g \approx$ 0.04 Hz at $\omega_{c}=$ 3 GHz). The dashed red line highlights a pessimistic room temperature limit for GHz resonator \textit{Q}'s. The parameters used here are $\gamma_{1}= 250$ Hz, $\gamma_{2}=\Delta_{ens}=1$ MHz. (b) Log-scale DC-sensitivities at 300 K for the MIT probe defined in Eq.3 for the same parameters.  The dashed red line highlights \textit{$Q_{s}$} = $10^5$. (c) Log-scale DC-sensitivities for \textit{$Q_{p}$} = $10^5$, measured using the point of maximum obtainable slope of an asymmetric Fano lineshape.}
\label{Fig4}
\end{figure*}

This limit can be further augmented through Fano-interference when considering scattering between the dressed state and another uncoupled resonant mode. Cobnsidering e.g. circular resonators, these possess a dispersion of whispering gallery modes whose phase will be mismatched depending on the relative mode number difference and spatial distribution. Scattering of the coupled mode can occur with any other mode order, either in the same cavity, or with an additional externally coupled cavity. If a dressed state is tuned to the frequency of a neighbouring mode by varying the coupling strength, or evanescently interfered with a mode of another resonator, Fano-type scattering is expected to occur, as well as magnetically induced transparency (MIT) when on-resonance \citep{Miroshnichenko2010}. An example of a detuning plot of a dressed state interfering with a $\pi$-mismatched mode is shown in Fig.3(a,i) with extracted ODMR spectrum plotted in Fig.3(a,ii). The resulting linewidth of the MIT is determined by the phase mismatch and the on-resonance overlap integral between the cavity mode and the dressed-state mode, and takes the approximate form: \begin{equation}
\Gamma_{MIT} \simeq \bigg\vert\frac{\Gamma_{p}\Gamma_{s}}{\Gamma_{p}+e^{-i\phi}\Gamma_{s}}\bigg\vert
\end{equation} where $\phi$ is the phase mismatch and $\Gamma_{s}$ is the scattering mode linewidth. This results in a spectral window which can possess a narrower linewidth then the dressed-state but which exhibits half its contrast, as plotted in Fig.3(a,ii). Alternatively, the asymmetric Fano line-shapes, which emerge with detuning, do not suffer from reduced contrast. The gradient of these slopes exceeds that of both the individual dressed state and the MIT window, as highlighted Fig.3(b).

Both bare and Fano-scattered dressed-states can be applied in schemes where sensing is conveyed through measured fluorescence contrast (see Appendix D). External magnetic fields can be inferred through the spin Larmor frequency shift in their presence via $\Delta\nu_{ESR} = \gamma_{e}B_{0}$, where $B_{0}$ is the projection of the magnetic field along the principal NV quantisation axis. Given that the dressed-state frequency is explicitly dependent on the detuning between the cavity resonance and the central frequency of the spin distribution, it is fair to assume that the dressed-state will shift with an identical gyromangetic ratio to that of an uncoupled NV spin. The frequency dependent fluorescence can be quantified in terms of the expected measured number of photons from an ensemble. During read-out, the ensemble is incoherently probed to saturation, thereby generating emission from both coupled and uncoupled spins is collected. Approximating the point of maximum slope of the resulting spectra as a function of $\Gamma$, sensitivity can be defined through:
\begin{equation}
\delta B_{ODMR} \simeq \frac{1}{\gamma_{e}}\frac{\Gamma}{C\sqrt{\zeta}}
\end{equation} 
where $C$ is the contrast of the peak height from the off-resonance intensity, and $\zeta$ is the shot-noise limited detected photon number for which $\zeta = \eta_{DetQe} \cdot \eta_{ext} \cdot N \cdot t_{m} \cdot \gamma_{cyc} \cdot \eta_{NVQE}$, where detector quantum efficiency $\eta_{DetQe}$ (70 $\%$), photon extraction efficiency $\eta_{ext} $ (50 $\%$), the number of photon emitters $N$, the read-out pulse duration time $t_{m}$ (1 s), the total cyclic transition from excited triplet-states through the singlet-states to the triplet ground-state ($m_s = 0$) $\gamma_{cyc}$ ($\sim130$ MHz), and the total quantum efficiency of the radiative decay rate from the $m_s = 0$ state $\eta_{NVQE}$ (80 $\%$) have been included. 

Fig.4 plots the predicted DC-field sensitivity for $\Gamma_{p}$ (calculated using the coupled modes quality \textit{$Q_{p}$}), $\Gamma_{MIT}$ (calculated using \textit{$Q_{p}$} and a scattering mode quality \textit{$Q_{s}$} = $10^5$), at 300 K, with an individual coupling strength calculated using a mode volume of 1 cm$^{3}$. The predicted sensitivity spans from nano to below atto T Hz$^{-\frac{1}{2}}$ but obtaining \textit{$Q_{p}$}  $ \geq10^6$  at room temperature is technically demanding, beyond which stabilisation of the resonance becomes challenging. However, even considering a pessimistic limit of \textit{$Q_{p}$}'s at $10^4$, sensitivities around aT Hz$^{-\frac{1}{2}}$  seem to be achievable, given the demonstrated achievable densities of $10^{18}$ cm$^{-3}$ \citep{Acosta2009}. Using the MIT window, the requirements are lessened, as seen in Fig.4(b) for a \textit{$Q_{s}$} of $10^5$. An almost linear independence of \textit{$Q_{p}$} is achieved up until \textit{$Q_{p}$} $ \geq$ \textit{$Q_{s}$}. Both plots highlight a decrease in sensitivity after a certain ensemble density before increasing further below aT Hz$^{-\frac{1}{2}}$, which highlights the transition into the cavity-protected regime. Further log-scale improvement is seen in Fig.4(c) when using the asymmetric Fano-lineshape when \textit{$Q_{s}$} $ \geq$ \textit{$Q_{p}$} = $10^{5}$ for the same parameters used in Fig.3(a), showing sensitivities approaching zT Hz$^{-\frac{1}{2}}$. 

In summary, the calculated dynamics highlight the manifestation of strong coupling and the diminishing influence of temperature for increasing NV ensemble sizes. Using a collective coupling strength and assuming that large ensembles will posses a dominantly Gaussian distribution, a temperature-dependent, steady-state, analytical expression for the cavity-protected dressed-state linewidth was derived, which was used to estimate ODMR based DC-field sensitivities. This predicted aT Hz$^{-\frac{1}{2}}$  sensitivity for densities approaching $10^{20}$ cm$^{-3}$, with further augmentation investigated through the possibility of scattering with external resonant modes showing zT Hz$^{-\frac{1}{2}}$ may be achievable. These results highlight the prospect of using both optical and spin degrees of freedom under ambient conditions to further explore cavity-QED and metrology schemes using NVs. 

\begin{acknowledgements}
We are grateful to Fedor Jelezko, Viatcheslav Dobrovitski, Dara McCutcheon and Jonas Neergaard-Nielsen for fruitful discussions and feedback. This work was supported by the Danish Innovation Foundation (EXMAD project).
\end{acknowledgements}

\appendix

\section{Appendix A: Expressions for $\langle\hat{\sigma}^{z}\rangle_{ss}$ and $\langle\hat{a}^{\dagger}\hat{a}\rangle_{ss}$ }

The primary intention here is to justify the Heisenberg-Langevin solution for the steady-state system output using the Holstein-Primakoff approximation, for a large ($N > 10^{5}$) ensemble coupled to a single cavity mode at room temperature. This allows for an analytical expression of the steady-state output , unattainable through direct numerical integration of the master equation or a Monte Carlo wave function approach, which accounts for system temperature and ensemble spectral dispersion shape (for $N$ two-level systems, the density matrix will possess a dimensionality of $[2^N \times 2^N]$ which becomes numerically intractable when $N > 10^{5}$ through both these procedures).

The system is first explored using a reduced master equation to assess and obtain an expression for the effect of temperature on the total inversion, as well as the steady-state number of remnant excitations, prior to populating the spin-ensemble states. It is assumed that the correlation time of the heat bath is much shorter then the system evolution time-scale as well as that the system and bath are initially uncorrelated. The Markov-approximation is therefore employed, in addition to assuming that all ensemble spins are identical. The Liouvillian operators described in the main text are defined as: 
\begin{eqnarray}
\Lagr_{\kappa(\bar{n})}[\rho_{s}]= &&\kappa(\bar{n}+1)(2\hat{a}\rho_{s}\hat{a}^{\dagger}-\{\hat{a}^{\dagger}\hat{a},\rho_{s}\}) \nonumber
 \\
 &&+\kappa\bar{n}(2\hat{a}^{\dagger}\rho_{s}\hat{a}-\{\hat{a}\hat{a}^{\dagger},\rho_{s}\})
\\\nonumber
\\\nonumber
\Lagr_{\gamma_{1}(\bar{n})}[\rho_{s}]= &&\frac{1}{2}[\gamma_{1}(\bar{n}+1)\sum_{k}^{N}(2\hat{\sigma}_{k}^{-}\rho_{s}\hat{\sigma}_{k}^{+}-\{\hat{\sigma}_{k}^{+}\hat{\sigma}_{k}^{-},\rho_{s}\})
\\
 &&+ \gamma_{1}\bar{n}\sum_{k}^{N}(2\hat{\sigma}_{k}^{+}\rho_{s}\hat{\sigma}_{k}^{-}-\{\hat{\sigma}_{k}^{-}\hat{\sigma}_{k}^{+},\rho_{s}\}) ]
\\\nonumber
\\
 \Lagr_{\gamma_{2}}[\rho_{s}]=&&\frac{\gamma_{2}}{2}\sum_{k}^{N}(\hat{\sigma}_{k}^{z}\rho_{s}\hat{\sigma}_{k}^{z}-\rho_{s})
\\\nonumber
\\ 
 \Lagr_{v}[\rho_{s}]=&&-\frac{v}{2}\sum_{k}^{N}(2\hat{\sigma}_{k}^{+}\rho_{s}\hat{\sigma}_{k}^{-}-\{\hat{\sigma}_{k}^{-}\hat{\sigma}_{k}^{+},\rho_{s}\})
\end{eqnarray}
Here the thermal excitation of the mode and ensemble is described through the temperature and frequency dependent average number of thermal photons: $\bar{n}(\omega,T) = 1/( e^{\frac{\hbar\omega}{k_{b}T}}-1)$. The operator expectation values are derived through cyclic permutation of the expectation values and reduced density matrix trace Tr$[\hat{O}\dot{\rho}_{s}]$. As the system is prepared and read-out using the incoherent probe, the total phase-invariance of the system can be assumed to remain unchanged for the instances considered here, namely the preparation stage before the coherent probe is applied. Under this assumption, and only considering correlations up to second order, the coupled rate equation reduces down to a set of four coupled equations from which the inversion and cavity occupation number form a closed pair of equations:
\begin{eqnarray}
\dot{\langle\hat{\sigma}^{z}\rangle}  =&&-(\gamma_{1}+v)\langle\hat{\sigma}^{z}\rangle(2\bar{n}+1)+(\gamma_{1}-v)\nonumber
\\
&&-i2g(\langle\hat{\sigma}^{+}\hat{a}\rangle-c.c.)
\\\nonumber
\\
\dot{\langle\hat{a}^{\dagger}\hat{a}\rangle}=&&2\kappa(\bar{n}-\langle\hat{a}^{\dagger}\hat{a}\rangle)-igN(\langle\hat{\sigma}^{-}\hat{a}^{\dagger}\rangle-c.c.)
\\\nonumber
\\
\dot{\langle\hat{\sigma}^{-}\hat{a}^{\dagger}\rangle}=&& -i(\tilde{\omega}_{c} + \tilde{\omega}_{k})\langle\hat{\sigma}^{-}\hat{a}^{\dagger}\rangle-\frac{1}{2}ig(\langle\hat{\sigma}^{z}\rangle+1)\nonumber
\\
&&-ig(N-1)\langle\hat{\sigma}^{+}\hat{\sigma}^{-}\rangle-ig\langle\hat{\sigma}^{z}\rangle\langle\hat{a}^{\dagger}\hat{a}\rangle
\\\nonumber
\\
\dot{\langle\hat{\sigma}^{+}\hat{\sigma}^{-}\rangle}= &&-ig\langle\hat{\sigma}^{z}\rangle(\langle\hat{\sigma}^{-}\hat{a}^{\dagger}\rangle-c.c.)\nonumber
\\
&&-(\gamma_{1} + v)\langle\hat{\sigma}^{+}\hat{\sigma}^{-}\rangle
\end{eqnarray}
where the tilde annotation signified complex frequencies which incorporates all the loss rates through the complex component. The dynamics extracted from these coupled equations directly highlight an ensembles collective polarisation and coherence for a given set of initial temperature and incoherent pump conditions, through the manifestation, or lack of, Rabi oscillations. The steady-state expressions for the inversion and cavity occupation number can then be extracted from the coupled equations and expressed as:
\begin{eqnarray}
\langle\hat{\sigma}^{z}\rangle_{ss}=&&\frac{(v-\gamma_{1})-(4\kappa/N)(\langle\hat{a}^{\dagger}\hat{a}\rangle_{in}-\bar{n})}{(v+\gamma_{1})(2\bar{n}+1)}
\\\nonumber
\\
\langle\hat{a}^{\dagger}\hat{a}\rangle_{ss} = &&\bar{n} - \frac{N(\gamma_{1}+v)}{4\kappa}\nonumber
\\
&&\times\bigg(\langle\hat{\sigma}^{z}\rangle_{in}(1+2\bar{n})-\frac{v-\gamma_{1}}{v+\gamma_{1}}\bigg)
\end{eqnarray}
where \textit{in} denotes the initial inversion and cavity photon number prior to incoherent pumping. Considering an initial cavity system in thermal equilibrium, $\langle\hat{\sigma}^{z}\rangle_{in}=0$ and $\langle\hat{a}^{\dagger}\hat{a}\rangle_{in}=\bar{n}$. 

\section{Appendix B: Total System Output}
Having outlined the effect of temperature and incoherent pump strength, the initial Hamiltonaina may be simplified within these approximations using expressions for the collective coupling strength, collective spin operators and the Holstein-Primakoff approximation:
\begin{eqnarray}
&&\sum_{k}^{N}\hat{\sigma}_{k}^{+} = \hat{S}^{+} = \hat{b}^{\dagger}\sqrt{1-\hat{b}^{\dagger}\hat{b}}\approx\hat{b}^{\dagger}
\\
&&\sum_{k}^{N}\hat{\sigma}_{k}^{z}=\hat{S}^{z}= 2\hat{b}^{\dagger}\hat{b}-1
\\
&&\Omega = \sum_{k}^{N}g_{k}
\end{eqnarray} 
The resulting Hamiltonian is defined as:
\begin{equation}
\hat{\mathcal{H}}_{HP} =\omega_{c}\hat{a}^{\dagger}\hat{a}  + \sum_{k}^{N}\omega_{k}\hat{b}^{\dagger}\hat{b} +\Omega(\hat{b}^{\dagger}\hat{a}-\hat{b}\hat{a}^{\dagger})
\end{equation}
The equations of motion for the expectation value of the two operators $\langle\hat{a}\rangle$ and $\langle\hat{b}\rangle$ can then be defined along with the input-output relations to derive an expression for the complex transmission of the cavity:
\begin{eqnarray}
\dot{\langle\hat{a}\rangle} =&&i[\hat{\mathcal{H}}_{HP},\langle\hat{a}\rangle]-\frac{\kappa}{2}\langle\hat{a}\rangle \nonumber
\\
=&&-\bigg(\frac{\kappa}{2}+i(\omega_{c}-\omega)\bigg)\langle\hat{a}\rangle+\Omega\langle\hat{b}\rangle -\eta_{in}\sqrt{\kappa/2} 
\\
\dot{\langle\hat{b}\rangle} =&&i[\hat{\mathcal{H}}_{HP},\langle\hat{b}\rangle]-\frac{\gamma_{1}}{2}\langle\hat{b}\rangle\nonumber
\\
=&&-\bigg(\frac{\gamma_1}{2}+i(\omega_{k}-\omega)\bigg)\langle\hat{b}\rangle-g_{k}\langle\hat{a}\rangle
\\
\eta_{tr}=&&\langle\hat{a}\rangle\sqrt{\kappa/2}
\\
\eta_{tr}=&&\eta_{re}-\eta_{in}
\end{eqnarray}
where $\eta_{tr}$ , $\eta_{re}$ and $\eta_{in}$  are the transmitted, reflected and injected external field amplitudes. Rearranging the steady-state expectation values for the total cavity transmission gives:
\begin{equation}
t_{c}(\omega)=\frac{\eta_{in}(i\kappa/2)}{\tilde{\omega}_{c}-\omega-\sum^{N}_{k}g_{k}^{2}/(\tilde{\omega}_{k}-\omega)}
\end{equation}
To account for the spectral dispersion, the number of coupled NV's are described as a continuous spectral density distribution $\rho(\omega)$ with a full width at half maximum $\Delta_{en}$ that is equivalent to the collective pure dephasing rate which includes both homogeneous and inhomogeneous contributions. The spectral density is initially expressed as:
\begin{equation}
\rho(\omega) = \sum^{N}_{k}g_{k}^{2}\delta(\omega-\omega_{k})/\Omega^{2}
\end{equation}
This is redefined in Eq.24 and simplified through the following substitution:
\begin{equation}
\frac{1}{(\tilde{\omega}_{k}-\omega)}=\int^{\infty}_{-\infty} d\omega'\frac{\delta(\omega'-{\omega}_{k})}{(\omega'-\omega-i\gamma_{1})}
\end{equation}
The integral in the denominator is then expanded using the Sokhotski-Plemlj theorem in the limit where $\gamma_{1}\ll\Delta_{en}$:
\begin{eqnarray}
&&\Omega^{2}\int^{\infty}_{-\infty}d\omega'\frac{\rho(\omega')}{(\omega-\omega')}
\\\nonumber
\\
&&\Rightarrow \Omega^{2}{\cal P}\int^{\infty}_{-\infty}d\omega'\frac{\rho(\omega')}{(\omega-\omega')}-i\pi\Omega^{2}\rho(\omega)
\end{eqnarray}
resulting in the transmission expression:
\begin{equation}
t_{c}(\omega) = \frac{i\kappa/2}{\Delta\omega+i(\kappa+\gamma_{1}+2\pi\Omega^{2}\rho(\omega,\Delta_{en}))-\Omega^{2}\Lambda}
\end{equation}
where $\Delta\omega = \omega_{c}-\omega$, $\rho(\omega)$ is defined in terms of the inhomogeneous broadening $\Delta_{ens}$, and $\Lambda$ is a principal value integral in Eq.28, interpreted as the resulting non-linear Lamb-shift, and the injected external field amplitude is set to unity. Finally assuming that all coupling strengths are equal, $\Omega^{2} = g^{2}N$, and may be further defined in terms of the temperature dependence of the number of spins occupying the same level. This way, the ordered subset of the steady-state thermally disordered spin ensemble can be accounted for, by defining the temperature dependence of $\langle\hat{\sigma}^{z}\rangle_{ss}$. This dependence is derived analytically from Eq.14, or can be extracted through fitting:
\begin{eqnarray}
\Omega_{T,N} &&=  g\sqrt{N\langle\hat{\sigma}^{z}(T)\rangle_{ss}} 
\\
&&= \gamma_{e}\sqrt{\frac{\hbar\omega_{c}\mu_{0}}{V_{\text{eff}}}}\sqrt{N\tanh\bigg(\frac{\omega_{0}}{4\pi k_{b}T}\bigg)}
\end{eqnarray}
where $\mu_{0}$ is the magnetic permeability, $\gamma_{e}$ is the spin gyromagnetic ratio and  $V_{\text{eff}}$ is the effective cavity mode volume. The resulting linewidth can be extracted from the transmission function by analysing the function poles which occur at $\omega = \Omega_{T,N}$ and is expressed as:
\begin{equation}
\Gamma_{p}=\frac{\kappa+\gamma_{1}+2\pi\Omega_{T,N}^2\rho(\Omega_{T,N},\Delta_{en})}{2}
\end{equation}       
The spectral dispersion may be defined using any distribution function; here we use Gaussian, Lorentzian and a pseudo-Voigt function to obtain analytical expressions that can be directly solved:
\begin{eqnarray}
&&\rho_{G}=\frac{\sqrt{\log   2}}{\Delta_{en}} e^{-(\frac{\Delta\omega}{\Delta_{en}})^2}
\\\nonumber
\\
&&\rho_{L}=\frac{\Delta_{en}}{2\pi((\Delta_{en}/2)^2+\Delta\omega^2)}
\\\nonumber
\\
&&\rho_{pv}=\vartheta\rho_L + (1-\vartheta)\rho_G, ~~~~~ 0\leq\vartheta\leq1
\end{eqnarray}
The total emission expected to be highlighted through ODMR will include both coupled and uncoupled spin states. This implies that the total transmission should be defined as the sum of the output of both steady state operators $\langle\hat{a}\rangle$ and $\langle\hat{b}\rangle$. The summed expression, plotted in Fig.2(a) in the main article is given as:
\begin{eqnarray}
I_{ODMR}(\omega) &&= \bigg\vert t_{c}(\omega)+t_{ens}(\omega)\bigg\vert^{2}
\\\nonumber
\\
&&=\bigg\vert t_{c}(\omega) +\frac{(\Delta\omega+i\kappa)\Lambda\sqrt{t_{c}(\omega)}}{\Omega_{T,N}^{2}}\bigg\vert^2
\end{eqnarray}

\section{Appendix C: Fano-Scattering}
Fano-interference can be summarised as the resultant spectrum of energy dissipating from an initial state to a final state through two competing paths that are energetically indistinguishable. This can be derived straightforwardly by defining the initial Hamiltonian for two different cavity modes with distinct coupling strengths to the same ensemble. In the scenario described here, this can be well-approximated by the overlap integral of competing emission profiles (which in this case are the dressed-state spectrum and the uncoupled mode spectrum), while accounting for their relative phase. To first-order approximation, the resulting spectrum can be expressed through:
\begin{equation}
I_{Fano}(\omega) = \vert t_{c}(\omega)+t_{ens}(\omega)+e^{-i\phi}t_{scatt}(\omega)\vert^{2 }
\end{equation}    
where $t_{scatt}(\omega)$ would typically be defined through a  Lorentzian profile, and $\phi$ is the relative phase difference for two competing modes. Looking at e.g. a circular disk resonator and exclusively analysing $1^{st}$ order radial modes, the relative phase mismatch of two different modes at frequencies $c/\lambda_{dress}$ and $c/\lambda_{scatt}$  can be expressed via:
\begin{equation}
\phi = \frac{\pi}{2}\bigg(\cos\bigg(\frac{2\pi rn}{\lambda_{dress}}\bigg)-\cos\bigg(\frac{2\pi rn}{\lambda_{scatt}}\bigg)\bigg)
\end{equation}    
where $r$ is the radius and $n$ is the refractive index of the disk resonator. 

\section{Appendix D: Magnetic Field Sensing}
A sensors measurement sensitivity to any external field, in this case magnetic, is fundamentally defined through:
\begin{equation}
\delta B_{min} =\sigma(t)\bigg/\frac{d\mathcal{S}}{dB} \approx \sqrt{\beta}/\gamma_{e}t_{m}\max \bigg\vert\frac{\partial \mathcal{I}}{\partial \nu_{ESR}}\bigg\vert
\end{equation}
where $\sigma(t)$ is the standard deviation of the sensors signal as a function of time and $dS/dB$ is the response of the system to a changing external magnetic field $B$. $\delta\mathcal{S}(t) \approx  \sqrt{\beta}$ is the corresponding shot noise of the total detected number of photons per measurement $\beta = \mathcal{I}t_{m}$. This is a general relation that can be applied to most measurement procedures with fairly intuitive and straightforward substitutions. 

For real measurements, the detected light is dispersed with a particular line shape and intensity that is set by the coherence time, scattering cross-section, and the total quantum efficiency of both excitation and detection of the polarised spin states. The point of maximum slope of this dispersion is where the largest shift in fluorescence will occur for a given change in external field, and can be approximately defined through the full width at half maximum of a symmetric dispersion through 
\begin{equation}
\max \bigg\vert\frac{\partial \mathcal{I}}{\partial \nu_{ESR}}\bigg\vert = \frac{C\mathcal{I}}{\Delta\nu \mathcal{P_{F}}}
\end{equation}
where $\mathcal{P_{F}}$ is a distribution function-dependent constant (for a Lorentzian and Gaussian lineshape this is equal to $4/3\sqrt{3}$ and $\sqrt{e/8 \text{ln} 2}$, respectively). This relation is used to assess the relative enhancement of the asymmetric Fano lineshapes in the main text. 

\bibliography{References}
\end{document}